\documentclass{ws-procs975x65}

\begin{document}

\title{CONFORMAL PHASE TRANSITION IN QCD LIKE THEORIES AND BEYOND}

\author{V. A. Miransky$^*$}

\address{Department of Applied Mathematics, University of Western Ontario,\\
London, Ontario N6A 5B7, Canada\\
$^*$E-mail: vmiransk@uwo.ca\\}

\begin{abstract}
The dynamics with an infrared stable fixed point
in the conformal window in QCD like theories
with a relatively large number of fermion flavors is reviewed.
The emphasis is on the description of a clear signature for the
conformal window, which in particular can be useful for lattice
computer simulations of these gauge theories.
\end{abstract}


\bodymatter

\section{Introduction}\label{sec1}

The Landau, or 
$\sigma$-model-like, phase transition \cite{Landau} is 
characterized by the following basic feature.
Around the critical point
$z=z_c$ (where $z$ is a generic notation for parameters of a theory,
as the coupling constant $\alpha$, number of particle flavors $N_f$, etc.),
an order parameter $X$ is
\begin{equation}
X = \Lambda f(z), \label{1}
\end{equation}
where $\Lambda$ is an ultraviolet cutoff and the function
$f(z)$ has such a non-essential singularity at $z=z_c$
that $\lim f(z)=0$ as $z$ goes to $z_c$ both in symmetric and
non-symmetric phases. The standard form for $f(z)$ is
$f(z) \sim (z - z_c)^{\nu}$, $\nu > 0$, around $z = z_c$ 
[for convenience, we assume that $z > z_c$ $(z < z_c)$ in the
nonsymmetric (symmetric) phase].\footnote {
Strictly speaking, Landau considered the mean-field
phase transition.
By the Landau phase transition, we
understand a more general class, when fields may have anomalous
dimensions \cite{W}.
}
The conformal phase transition (CPhT), whose conception was introduced 
in Ref. \refcite{cpt}, 
is a very different continuous phase transition. It is
defined as a phase transition in which an order parameter $X$ is given by
Eq. (\ref{1})
where $f(z)$ has such an {\it essential} 
singularity
at $z=z_c$ that while
\begin{equation}
\lim_{z \to z_c} f(z) =0 \label{2}
\end{equation}
as $z$ goes to $z_c$ from the side of the non-symmetric phase,
$\lim f(z) \ne 0 $
as $z \to z_c$ from the side of the symmetric phase
(where $X \equiv 0$).
Notice that since the relation (\ref{2}) ensures that
the order parameter $X \to 0$ as $z \to z_c$,
the phase transition is continuous.

There are the following basic differences between the Landau
phase transition (LPhT) and the CPhT one \cite{cpt}:

\begin{enumerate}

\item 
In the case of the LPhT, 
masses of light excitations are continuous functions
of the parameters $z$ around the critical point $z=z_c$
(though they are non-analytic at $z=z_c$). 
In the case of the CPhT, the situation is different:
there is an abrupt change of the spectrum of light excitations,
as the critical point $z=z_c$ is crossed. This implies that 
the effective actions 
describing low energy dynamics in the phases with $z < z_c$ and
$z > z_c$ are different in a system with CPhT. 

\item
Unlike the LPhT, the parameter $z$ governing
the CPhT is connected with a marginal
operator [in the LPhT phase transition, such a parameter
is connected with a relevant operator; it is usually a
mass term].

\item 
The fact that the parameter $z$ is connected with a
marginal operator in the CPhT implies that in the continuum
limit, when $z \to z_c + 0$, the conformal symmetry is
broken by a marginal operator in nonsymmetric phase, i.e.,
there is a conformal anomaly. 

\item Unlike the LPhT, in the case of CPhT, 
the structures of renormalizations (i.e., the renormalization
group at high momenta) are different in symmetric phase and 
nonsymmetric one.

\end{enumerate}

In relativistic field theory, the CPhT is realized in the two
dimensional Gross-Neveu (GN) model \cite{GN} at the
critical coupling constant $g_c = 0$, reduced (or defect)
QED \cite{GGM,Rey,K}, and quenched QED \cite{QED1,QED2,QED3,QED4}. 
It was suggested that the chiral phase transition with respect to
the number of fermion flavors $N_f$ in QCD is a CPhT one 
\cite{ATW,cpt}.
In condensed matter physics, a CPhT like phase transition is realized
in the Berezinskii-Kosterlitz-Thouless (BKT)
model \cite{BKT} and, possibly, graphene \cite{GGMS}. 

Recently, the interest to the dynamics with the CPht phase transition 
has essentially increased. It is in particular
connected with a progress in numerical
lattice studies of gauge theories with a varied number of fermion flavors
(for a recent review, see 
Ref. \refcite{EP}), the revival of the interest to the
electroweak symmetry breaking based on the walking technicolor 
like dynamics \cite{wt1,wt2} (for a recent review, see Ref. \cite{fs}), and 
intensive studies of graphene, a single atomic layer of graphite
(for a review, see Ref. \refcite{graphene}).

\section{Dynamics in the conformal window in QCD-like theories}\label{sec2}

\subsection{General description}\label{2a}

In this section, we will consider the problem of the existence of
a nontrivial conformal dynamics in 3+1 dimensional
non-supersymmetric vector like gauge theories, with
a relatively large number of fermion flavors $N_f$. We will discuss
their phase diagram in the
($\alpha^{(0)}, N_f$) plane, where $\alpha^{(0)}$ is
the bare coupling constant. We also discuss
a clear signature for the conformal window in lattice computer simulations 
of these theories suggested quite time ago in Ref. \refcite{VM}. 

The roots of this problem go back to a work of Banks and Zaks \cite{BZ}
who were first to discuss the consequences of the existence of
an infrared-stable fixed point $\alpha=\alpha^{*}$ for $N_f>N_f^{*}$ in
vector-like gauge theories \cite{J}.
The value $N_f^{*}$ depends on the gauge group:
in the case of SU(3) gauge group, $N_f^{*}=8$ in the two-loop approximation.
In Nineties, a new insight in this problem \cite{ATW,cpt} was, on the one hand, connected
with using the results of the analysis of the Schwinger-Dyson (SD) equations
describing chiral symmetry breaking in quenched QED \cite{QED1,QED2,QED3,QED4}
and, on the other hand, with the discovery of the conformal window in $N=1$
supersymmetric QCD \cite{S}.

In particular, Appelquist, Terning, and  Wijewardhana \cite{ATW}
suggested that,
in the case of the gauge group SU($N_c$),
the critical value $N_f^{cr}\simeq4N_c$
separates a phase with no confinement and chiral symmetry breaking
($N_f>N_f^{cr}$) and a phase with confinement and with chiral symmetry
breaking ($N_f<N_f^{cr}$).  The basic point for this suggestion was
the observation that at $N_f>N_f^{cr}$ the value of the infrared fixed
point $\alpha^{*}$ is smaller than a critical value
$\alpha_{cr}\simeq\frac{2N_c}{N_c^2-1}\frac{\pi}{3}$,
presumably needed to generate the chiral condensate \cite{QED1,QED2,QED3,QED4}.

The authors of Ref. \refcite{ATW} considered only the case when the running coupling
constant $\alpha(\mu)$ is less than the fixed point $\alpha^{*}$.
In this case the dynamics is asymptotically free (at short distances)
both at $N_f<N_f^{cr}$ and
$N_f^{cr}<N_f<N_f^{**} \equiv\frac{11N_c}{2}$.
Yamawaki and the author \cite{cpt} analyzed the dynamics in the whole
($\alpha^{(0)}, N_f$) plane and suggested the ($\alpha^{(0)}, N_f$)-phase diagram of
the SU($N_c$) theory, where $\alpha^{(0)}$ is 
the bare coupling constant (see Fig \ref{fig1} below).\footnote{This phase diagram
is different from the original
Banks-Zaks diagram \cite{BZ}.} 
In particular, it was pointed out that one can get an interesting
non-asymptotically free dynamics when the bare coupling
constant $\alpha^{(0)}$ is {\it larger} than $\alpha^{*}$,
though not very large. 

The dynamics with $\alpha^{(0)}>\alpha^*$ admits a continuum
limit and is interesting in itself.
Also, its better understanding can be important for establishing
the conformal window in lattice computer simulations of
the SU($N_c$) theory with such large values of $N_f$.
In order to illustrate this, let us consider the following example.
For $N_c=3$ and $N_f=16$, the value of the infrared fixed point $\alpha^*$
calculated in the two-loop approximation is small:
$\alpha^*\simeq$0.04.  To reach the asymptotically free
phase, one needs to take the bare coupling $\alpha^{(0)}$ less than this
value of $\alpha^*$.
However, because of large finite size effects, the lattice
computer simulations of the SU(3) theory with such a small $\alpha^{(0)}$
would be unreliable.
Therefore, in this case, it is necessary to consider the dynamics with
$\alpha(\mu)>\alpha^*$.

In Ref. \refcite{VM}, this author suggested a clear signature of
the existence of the infrared fixed point $\alpha^*$,
which in particular can be useful for lattice computer simulations.
The signature is based on two characteristic features of the  
the spectrum of low energy excitations
in the presence of a bare fermion mass in the conformal window:
a) a strong (and simple) dependence of the masses of all the colorless 
bound states (including glueballs) 
on the bare fermion mass,
and b) unlike QCD with a small $N_f$ ($N_f$=2 or 3), 
glueballs are lighter than bound states composed of fermions, 
if the value of the infrared fixed point is not too large.

\subsection{Phase diagram}
\label{2b}

\begin{figure}[htbp]
\begin{center}
\epsfxsize=8cm
\ \epsfbox{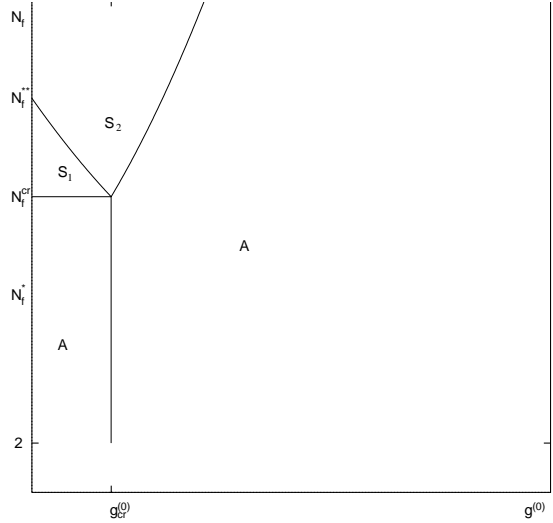}
\end{center}
\caption[]{The phase diagram in an SU($N_c$) gauge model. The
coupling constant $g^{(0)}=\sqrt{4\pi\alpha^{(0)}}$ and $S$ and
$A$ denote symmetric and asymmetric phases, respectively.}
\label{fig1}
\end{figure}

The phase diagram in the ($\alpha^{(0)}, N_f$)-plane in the
$SU(N_c)$ gauge theory is shown in Fig. \ref{fig1}.
The left-hand portion of the curve in this figure coincides with
the line of the infrared-stable fixed points $\alpha^*(N_f)$ \cite{J}:
\begin{equation}
\alpha^{(0)} = \alpha^* = - \frac{b}{c},
\label{alpha^*}
\end{equation}
where
\begin{eqnarray}
b&=&\frac{1}{6\pi} (11N_c - 2N_f),
\label{b} \\
c&=&\frac{1}{24\pi^2} (34N_c^2 - 10N_cN_f - 3\frac{N_c^2 -
1}{N_c}N_f).
\label{c}
\end{eqnarray}
It separates two symmetric phases, $S_1$ and $S_2$,
with $\alpha^{(0)}<\alpha^*$ and $\alpha^{(0)}>\alpha*$,
respectively.  Its lower end is $N_f=N_f^{cr}$ (with
$N_f^{cr}\simeq 4N_c$ if $\alpha_{cr}\simeq\frac{2N_c}{N_c^2-1}\frac{\pi}{3}$):
at $N_f^*<N_f<N_f^{cr}$ the infrared fixed point is washed out
by generating a dynamical fermion mass (here $N_f^*$ is the value
of $N_f$ at which the coefficient $c$ in Eq. (\ref{c}) 
becomes positive and the fixed point disappears). 

The horizontal, $N_f=N_f^{cr}$, line describes a phase transition
between the symmetric phase $S_1$ and the phase with confinement
and chiral symmetry breaking.
As it was suggested in Ref.\refcite{ATW}, based on a similarity of
this phase transition with that in quenched $QED_4$ \cite{QED2,QED3},
there is the following scaling law for $m^2_{dyn}$:
\begin{equation}
m^2_{dyn}\sim \Lambda^2_{cr}\exp\left(-\frac{C}
{\sqrt{\frac{\alpha^*(N_f)}{\alpha_{cr}} - 1}}\right),
\label{mdyn}
\end{equation}
where the constant $C$ is of order one and $\Lambda_{cr}$ is a scale
at which the running coupling is of order $\alpha_{cr}$.
It is a CPhT phase transition with an essential singularity
at $N_f=N_f^{cr}$.

At last, the right-hand portion of the curve on the diagram occurs
because at large enough values of the bare coupling, spontaneous
chiral symmetry breaking takes place for any number $N_f$ of
fermion flavors.  This portion describes a phase transition called
a bulk phase transition in the literature, and it is presumably
a first order phase transition.
\footnote{The fact that spontaneous chiral symmetry breaking takes
          place for any number of fermion flavors, if $\alpha^{(0)}$
          is large enough, is valid at least for lattice theories
          with Kogut-Susskind fermions.
          Notice however that since the bulk phase transition is a
          lattice artifact, the form of this portion of the curve
          can depend on the type of fermions used in simulations.}
The vertical line ends above $N_f$=0 since in pure gluodynamics
there is apparently no phase transition between weak-coupling and
strong-coupling phases.

\subsection{Signature for the conformal window}
\label{2c}

Up to now we have considered the case of a chiral invariant action.
But how will the dynamics change if a bare fermion mass term is
added in the action?
This question is in particular relevant for lattice computer
simulations:  for studying a chiral phase transition on a finite
lattice, it is necessary to introduce a bare fermion mass.
As was pointed out in Ref. \cite{VM}, adding even an arbitrary small bare fermion
mass results in a dramatic changing the dynamics both in the
$S_1$ and $S_2$ phases.

Recall that in the case of confinement SU($N_c$) theories,
with a small, $N_f<N_f^{cr}$, number of fermion flavors,
the role of a bare fermion mass $m^{(0)}$ is minor if
$m^{(0)}<<\Lambda_{QCD}$ (where $\Lambda_{QCD}$ is a confinement
scale).  The only relevant consequence is that massless
Nambu-Goldstone pseudoscalars get a small mass (the PCAC dynamics).

The reason for that is the fact that the scale $\Lambda_{QCD}$,
connected with a conformal anomaly, describes the breakdown of the
conformal symmetry connected {\it both} with perturbative
and nonperturbative dynamics:  the running coupling and the
formation of bound state.
Certainly, a small bare mass $m^{(0)}<<\Lambda_{QCD}$ is
irrelevant for the dynamics of those bound states.

Now let us turn to the phases $S_1$ and $S_2$, with $N_f>N_f^{cr}$.
There
is still the conformal anomaly in these phases:  because of the running of
the effective coupling constant, the conformal symmetry is broken.
It is restored only if $\alpha^{(0)}$  is equal to the
infrared fixed point $\alpha^{*}$.
However, the essential difference with respect to confinement
theories is that this conformal anomaly 
have nothing to do with the dynamics forming bound states:
Since at $N_f>N_f^{cr}$ the effective coupling is relatively weak,
it is impossible to form bound states from $\it{massless}$
fermions and gluons (recall that the $S_1$ and $S_2$ phases are chiral
invariant).

Therefore the absence of a mass for fermions and gluons is a key
point for {\it not} creating bound states in those phases.
The situation changes dramatically if a bare fermion mass is
introduced:  indeed, even weak gauge, Coulomb-like, interactions
can easily produce bound states composed of massive constituents,
as it happens, for example, in QED, where electron-positron
(positronium) bound states are present.
To be concrete, let us consider the case when all fermions
have the same bare mass $m^{(0)}$.
It leads to a mass function $m(q^2)\equiv{B(q^2)/A(q^2)}$ in the fermion
propagator $G(q)=(\hat{q}A(q^2)-B(q^2))^{-1}$.
The current fermion mass $m$ is given by the relation
\begin{equation}
m(q^2)\vert_{q^2=m^2}=m.
\label{m}
\end{equation}

For the clearest exposition, let us consider a particular theory
with a finite cutoff $\Lambda$ and the bare coupling constant
$\alpha^{(0)}=\alpha(q)\vert_{q=\Lambda}$
being not far away from the fixed point $\alpha^*$.
Then, the mass function is changing in the ``walking'' regime \cite{wt2} with
$\alpha(q^2)\simeq\alpha^*$.  It is
\begin{equation}
m(q^2)\simeq m^{(0)}\left(\frac{M}{q}\right)^{\gamma_m}
\label{m(q)}
\end{equation}
where 
$\gamma_m$ is the anomalous dimension of the operator $\bar{\psi}\psi$:
$\gamma_m = 3 - d_{\bar{\psi}\psi}$ with $d_{\bar{\psi}\psi}$ being the
dynamical dimension of this operator. In the walking regime,
$\gamma_m\simeq1-(1-\frac{\alpha^*}{\alpha_{cr}})^{1/2}$ 
(see Refs. \refcite{QED3,wt2}).

Eqs.(\ref{m}) and (\ref{m(q)}) imply that
\begin{equation}
m\simeq \Lambda\left(\frac{m^{(0)}}{\Lambda}\right)^{\frac{1}
{1 + \gamma_m}}.
\label{m1}
\end{equation}
Recall that the anomalous dimension $\gamma_m \geq 0$, and 
$\gamma_m \lesssim 2$ in the ``walking'' regime.

There are two main consequences of the presence of the bare mass:

(a) bound states, composed of fermions, occur in the spectrum
of the theory.  The mass of a n-body bound state is
$M^{(n)}\simeq{nm}$. Therefore they satisfy the scaling
\begin{equation}
M^{(n)}\simeq{nm} \sim n \left(m^{(0)}\right)^{\frac{1}{1 + \gamma_m}}.
\label{scaling}
\end{equation}
(b) At low momenta, $q< m$, fermions and their bound states decouple.
There is a pure SU($N_c$) Yang-Mills theory with confinement.
Its spectrum contains glueballs.

To estimate glueball masses, notice that at momenta $q< m$, the
running of the coupling is defined by the parameter $\bar{b}$
of the Yang-Mills theory,
\begin{equation}
\bar{b}= \frac{11}{6\pi}N_c.
\label{barb}
\end{equation}
Therefore the glueball masses $M_{gl}$ are of order
\begin{equation}
\Lambda_{YM}\simeq m\exp(-\frac{1}{\bar{b}\alpha^*}).
\label{YM}
\end{equation}

For $N_c=3$, we find from Eqs.(\ref{b}), (\ref{c}), and
(\ref{barb}) that
$\exp(-\frac{1}{\bar{b}\alpha^*})$
is $6\times{10^{-7}}$, $2\times{10^{-2}}$, $10^{-1}$,
and $3\times{10^{-1}}$ for
$N_f$=16, 15, 14, and 13, respectively.
Therefore at $N_f$=16, 15 and 14, the glueball masses are
essentially lighter than the masses of the bound states composed of
fermions. 

The situation is similar to that in confinement QCD with heavy 
(nonrelativistic) quarks,
$m>>\Lambda_{QCD}$.
However, there is now a new important point.
In the conformal window,
{\it any} value of $m^{(0)}$ (and therefore $m$) is ``heavy'':
the fermion mass $m$ sets a new scale in the theory, and the
confinement scale $\Lambda_{YM}$ (\ref{YM}) is less, and rather often
much less, than this scale $m$. One could say that the latter plays a
role of a dynamical ultraviolet cutoff for the pure YM theory.

This leads to a spectacular ``experimental'' signature of the
conformal window in lattice computer simulations: the masses of all
colorless bound states, including
glueballs, decrease as
$(m^{(0)})^{\frac{1}{1+\gamma_m}}$
with the bare fermion mass $m^{(0)}$ for {\it all}
values of $m^{(0)}$ less than cutoff $\Lambda$. Moreover, one should expect
that glueball masses are lighter than the masses of the bound states composed
of quarks.

Few comments are in order:

(1) The phases $S_1$ and $S_2$ have essentially the same
long distance dynamics.
They are distinguished only by their dynamics at short distances:
while the dynamics of the phase $S_1$ is asymptotically free,
that of the phase $S_2$ is not. Also, while around the infrared fixed point
$\alpha^{*}$ 
the sign of the beta function is negative in $S_1$, 
it is positive in $S_2$.\cite{cpt}.
When all fermions are massive (with the current mass $m$),
the continuum limit $\Lambda\rightarrow\infty$ of the $S_2$-theory is a
non-asymptotically free confinement theory.
Its spectrum includes colorless bound states composed of fermions
and gluons.
For $q<m$ the running coupling $\alpha(q)$ is the same as in pure
SU($N_c$) Yang-Mills theory, and for all $q \gg m$
$\alpha(q)$ is very close to
$\alpha^*$ (``walking'', actually, ``standing'' dynamics).
For those values $N_f$ for which $\alpha^*$ is small
(as $N_f$=16, 15 and 14 at $N_c$=3), glueballs are much lighter than
the bound states composed of fermions.
Notice that unlike the case with $m=0$, corresponding to the unparticle
dynamics \cite{G}, there exists a conventional S-matrix
in this theory.

(2) In order to get the clearest exposition, we assumed such estimates as
$N_f^{cr}\simeq 4N_c$ for $N_f^{cr}$ and
$\gamma_m=1-{\sqrt{1-\frac{\alpha^*}{\alpha_{cr}}}}$
for the anomalous dimension $\gamma_m$. While the latter
should be reasonable for $\alpha^*<\alpha_{cr}$ (and especially for
$\alpha^*<<\alpha_{cr}$) \cite{QED3}, the former is based on the
assumption that $\alpha_{cr}\simeq\frac{2N_c}{N_c^2 - 1}\frac{\pi}{3}$
which, though seems reasonable, might be crude for some values
of $N_c$.  It is clear however that the dynamical picture presented
above is essentially independent of those assumptions.

\subsection{Lattice computer simulations}
\label{2d}

During last two years, there has been an essential progress in the
lattice computer simulations of gauge theories with a varied number of
fermion flavors.\footnote{For pioneer papers in this area, see
Refs. \refcite{Br,Iw,DM}.} For a recent review, see 
Ref. \refcite{EP} and the
papers of Tom Appelquist, George Fleming, Kieran Holland, Julius Kuti, 
Maria Lombardo, and Donald Sinclair in this volume.  

This author is certainly not an expert in lattice computer simulations.
Here I would like to discuss this topic only in the connection with the
phase diagram and the signature of the conformal window considered in the
Secs. \ref{2b} and \ref{2c} above.

In Ref. \refcite{ML}, based on the fact that the sign of the beta function
changes from negative to positive when the line between the $S_1$ 
and $S_2$ phases is crossed, the existence of the conformal window in
QCD with $N_f = 12$ was studied by using the measurements of the chiral
condensate and the mass spectrum. The analysis supports the existence of
the conformal window in this theory.

In Ref. \refcite{DGH}, the scaling law (\ref{scaling}) was 
rediscovered and applied to the study of the conformal window in
the SU(3) lattice gauge theory with two flavors of color sextet
fermions (the parameter $y_m$ in Ref. \refcite{DGH} is connected with
the anomalous dimension $\gamma_m$ as $y_m = 1 + \gamma_m$). The main
conclusion of that study was that $y_m \sim 1.5$ 
($\gamma_m \sim 0.5$).
This value is smaller than $\gamma_m \simeq 1$ in walking technicolor
and at this moment it is unclear whether this theory contains an 
infrared-stable fixed point.

The authors of Ref. \refcite{LDD} studied the spectrum of mesons and glueballs
in the SU(2) lattice gauge theory with adjoint fermions. They found that
for light constituent fermions the lightest glueballs are lighter than the
lightest mesons. It is tempting to speculate that in accordance with
the signature for the conformal window discussed above, this fact indicates
on the existence of a infrared fixed point in this theory. However, as 
the authors point out, a lot of issues should still be clarified in order
to reach a solid conclusion.  

It is clear that lattice simulations of gauge theories with varied numbers of
fermion flavors are crucial  
for further progress in our understanding of such dynamics. The important point
is that CPhT is a long range interactions phenomenon, which is very sensitive
to any screening and finite-size effects. The progress made in this area during
last two years is certainly encouraging.

\section{Final comments}
\label{3}

At present the interest to the dynamics with an infrared stable fixed point 
in the conformal window and related issues, such as the conformal phase 
transition, is quite high. In such a brief review, it would be impossible
to describe all the recent developments in this area. Let me just mention  
the following ones, intimately related to the issues discussed above:

a) The holographic (gauge/gravity duality) approach to describe the conformal
phase transition have been recently considered in Refs. \refcite{Rey,K}.

b) Nonperturbative approaches to the calculation of the beta function in QCD
with different numbers of fermion flavors $N_f$ were developed in Refs. 
\refcite{Ch,Ryt}.

c) QCD with different $N_f$ was studied by using renormalization group flaw
equations in Ref. \refcite{Gies}.

d) The problem of the existence of the conformal window in QCD on 
${\bf R^{3}} \times {\bf S^{1}}$ has been analyzed in Ref. \refcite{Pop}. 

e) Lattice computer simulations of the phase diagram of graphene have been
recently realized in Refs. \refcite{Drut,Hands}.

\section{Acknowledgments}

I am grateful to the organizers of SCGT09 Workshop, in particular Koichi
Yamawaki, for their warm hospitality. This work was supported by the
Natural Sciences and Engineering Research Council of Canada.

\end{document}